# Evidence of magnetoelectronic electromagnon mediated transport in flexoelectronic heterostructures

Anand Katailiha[1‡], Paul C. Lou[1‡], Ravindra G. Bhardwaj[1‡], Ward P. Beyermann[2], and Sandeep Kumar[1,3,*]

[1] Department of Mechanical Engineering, University of California, Riverside, CA 92521, USA

[2] Department of Physics and Astronomy, University of California, Riverside, CA 92521, USA

[3] Materials Science and Engineering Program, University of California, Riverside, CA 92521, USA

[*] Corresponding author

[‡] Equal contribution

Email: skumar@engr.ucr.edu

**Abstract**

The superposition of atomic vibrations and flexoelectronic effect gives rise to a cross correlation between free charge carriers and temporal magnetic moment of phonons in conducting heterostructures under an applied strain gradient. The resulting dynamical coupling is expected to give rise to quasiparticle excitations called as magnetoelectronic electromagnon that carries electronic charge and temporal magnetic moment. Here, we report experimental evidence of magnetoelectronic electromagnon in the freestanding degenerately doped p-Si based heterostructure thin film samples. These quasiparticle excitations give rise to long-distance (>100μm) spin transport; demonstrated using spatially modulated transverse magneto-thermoelectric and non-local resistance measurements. The magnetoelectronic electromagnons are non-reciprocal and give rise to large magnetochiral anisotropy ($0.352 \ A^{-1}T^{-1}$) that diminishes at lower temperatures. The superposition of non-reciprocal magnetoelectronic electromagnons gives rise to longitudinal and transverse modulations in charge carrier density, spin density and magnetic moment; demonstrated using the Hall effect and edge dependent magnetoresistance measurements, which can also be called as inhomogeneous magnetoelectronic multiferroic effect. These quasiparticle excitations are analogues to photons where time dependent polarization and temporal magnetic moment replaces electric and magnetic field, respectively and most likely topological because it manifests topological Nernst effect. Hence, the magnetoelectronic electromagnon can potentially give rise to quantum interference and entanglement effects in conducting solid state system at room temperature in addition to efficient spin transport.

## I. Introduction

An electromagnon[1-3] is a quasiparticle in a solid-state system that can be considered analogues to an electromagnetic wave or photon, where time dependent electric polarization and magnetic moment replace electric and magnetic fields, respectively, as shown in Figure 1 (a). A type of electromagnon, magneto active phonon, can also be considered in the framework of dynamical multiferroicity[4,5], where the superposition of time dependent polarization of phonon ($\partial_t \boldsymbol{P}$) and ferroelectric polarization($\boldsymbol{P}_{FE}$) in ferroelectric materials gives rise to temporal magnetic moment[6] ($\boldsymbol{M}_t \propto \boldsymbol{P}_{FE} \times \partial_t \boldsymbol{P}$). Recently, an electronic dynamical multiferroicity[7] ($\boldsymbol{M}_t \propto \boldsymbol{P}_{F-El} \times \partial_t \boldsymbol{P}$) was experimentally reported in the metal/oxide/degenerately doped Si thin film samples, where flexoelectronic polarization ($\boldsymbol{P}_{F-El}$) replaces the ferroelectric polarization. The electronic dynamical multiferroicity may induce an electromagnon like excitations in the conducting and non-ferroelectric materials. Such a discovery can potentially change the current scientific understanding of the electronic systems with application to quantum computing, spintronics, thermoelectrics and other future quantum devices.

The flexoelectronic effect is essential for electronic dynamical multiferroicity. In a metal/semiconductor (doped) heterostructure, it can be defined as an electronic response to an applied strain gradient as shown in Figure 1 (b,c)[8]. The applied strain gradient give rise to interfacial flexoelectric effect[9], gradient in the band-structure[10], band gap and charge carrier mobility in the bulk of the semiconductor. Tian et al. [9], recently, experimentally demonstrated a large piezoelectric like response in the Si interface under bending loads, which we interpret as interfacial flexoelectric effect in this study. Similarly, Wang et al. [10] reported a large out of plane polarization (~0.03 Cm$^{-2}$) in Si due to strain gradient. It, then, triggers a charge carrier injection/extraction from the metal layer to the doped semiconductor layer depending upon the sign (slope) of the strain gradient as shown in Figure 1 (b,c)[8]. The charge carrier concentration of electrons in a typical metal is of the order of $10^{22}$ cm$^{-3}$. Whereas, the charge carrier concentration in a typical degenerately doped semiconductor is $10^{19}$ cm$^{-3}$. Due to flexoelectronic effect, the charge carrier concentration is semiconductor layer (p-Si in present study) can rise to the order of $10^{21}$ cm$^{-3}$, which is called as flexoelectronic doping as demonstrated recently[8]. The gradient of excess charge carrier concentration ($n'$) give rise to a flexoelectronic polarization ($P_{F-El} \propto n'$) in the semiconductor layer as shown in Figure 1 (c)[8]. Such heterostructure samples under applied strain gradient are called as flexoelectronic heterostructures by us. Hence, the electromagnon like excitations in metal/doped semiconductor thin film heterostructure samples can be described as:

$$M_t^{\pm <hlk>} \propto P_{F-El} \times \partial_t P^{\pm} \propto n' \times \partial_t P^{\pm} \tag{1}$$

We call these excitations in electronic systems as magnetoelectronic electromagnon since they couple to electronic charge as opposed to ionic charge in case of magnetoelectric electromagnon observed in insulating multiferroics. In this study, we present experimental evidence of non-reciprocal magnetoelectronic electromagnon in flexoelectronically doped p-Si based heterostructure thin films samples at room temperature. We use spatially modulated magneto-thermoelectric and non-local resistance measurements to uncover the long-distance spin transport behavior from magnetoelectronic electromagnons. The non-reciprocity of magnetoelectronic electromagnons is discovered using non-reciprocal transport measurements. Using Hall effect and edge dependent magnetoresistance measurements, we also discover spatial modulations in charge carrier concentration, spin density and magnetic moment due to the superposition of the topological magnetoelectronic electromagnons, which can be considered as inhomogeneous magnetoelectronic multiferroic effect.

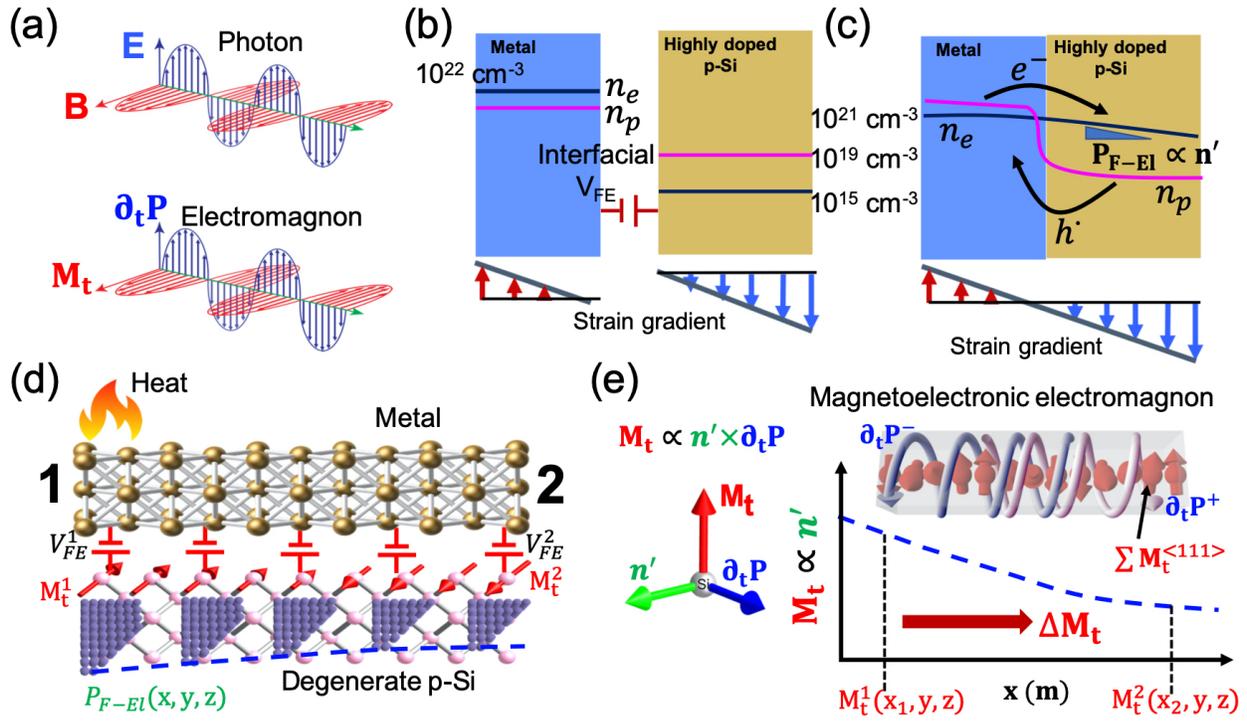

Figure 1. (a) A schematic showing that electromagnon are analogues to photon. (b) An approximate distribution of charge carrier concentration in a metal/p-Si (degenerate) heterostructure. (c) A schematic showing the flexoelectronic charge carrier injection in doped Si from metal layer due to interfacial flexoelectric effect leading to gradient of charge carrier in Si and flexoelectronic polarization. (d) A schematic showing the perturbation of the flexoelectronic charge carrier gradient due to an application of the heat at location 1 and subsequent decay to location 2. (e) A schematic showing the expected longitudinal decay of flexoelectronic charge carrier gradient leading to generation of magnetoelectronic electromagnon that carries electronic charge ($\Delta n$) and magnetic

moment ($\Delta M_t$). The superposition of magnetoelectronic electromagnons spatial modulations in spin angular momentum.

## II.    Experimental results

Traditionally, electromagnon[1-3] and other quasiparticle excitations are studied using spectroscopic techniques. However, we rely on transport measurements to uncover the evidence of magnetoelectronic electromagnon. We hypothesized an experimental scheme having a freestanding Si based flexoelectronic heterostructure where strain gradient from residual stresses will give rise to uniform flexoelectronic effect along the length of the sample. However, when the equilibrium state is modified at any given location in the heterostructure (using heat at location 1 as shown in Figure 1 (d)), the interfacial flexoelectric effect ($V_{FE}^1$) and consequently flexoelectronic polarization of the Si layer will change at this location 1 as shown in Figure 1 (d). Whereas the interfacial flexoelectric effect (($V_{FE}^2$) and flexoelectronic polarization at distance far away from the heat source remains same as shown in Figure 1 (d). As a result, the temporal magnetic moment ($\boldsymbol{M}_t^1$) at location 1 will be significantly larger than at location 2 ($\boldsymbol{M}_t^2$) as shown in Figure 1 (d,e). A longitudinal gradient in the temporal magnetic moment ($\Delta M_t$) will arise across the macroscopic sample in addition to the temperature gradient as shown in Figure 1 (d,e). A disturbance wave will be generated that will carry the non-equilibrium temporal magnetic moment (or spin angular momentum) along with the heat as shown in Figure 1 (d,e). As a consequence, magnetoelectronic electromagnons excitations will transport electronic charge ($\Delta n$) as well as magnetic moment ($\Delta M_t$) across the macroscopic sample as shown in Figure 1 (d,e). Therefore, the transport behavior from magnetoelectronic electromagnons can be studied using magneto transport measurements. It is noted that the spin injection from Py to Si layer will be there but it will not be the primary driver of the longitudinal spin chemical potential.

## A. Magneto-thermoelectric transport measurement

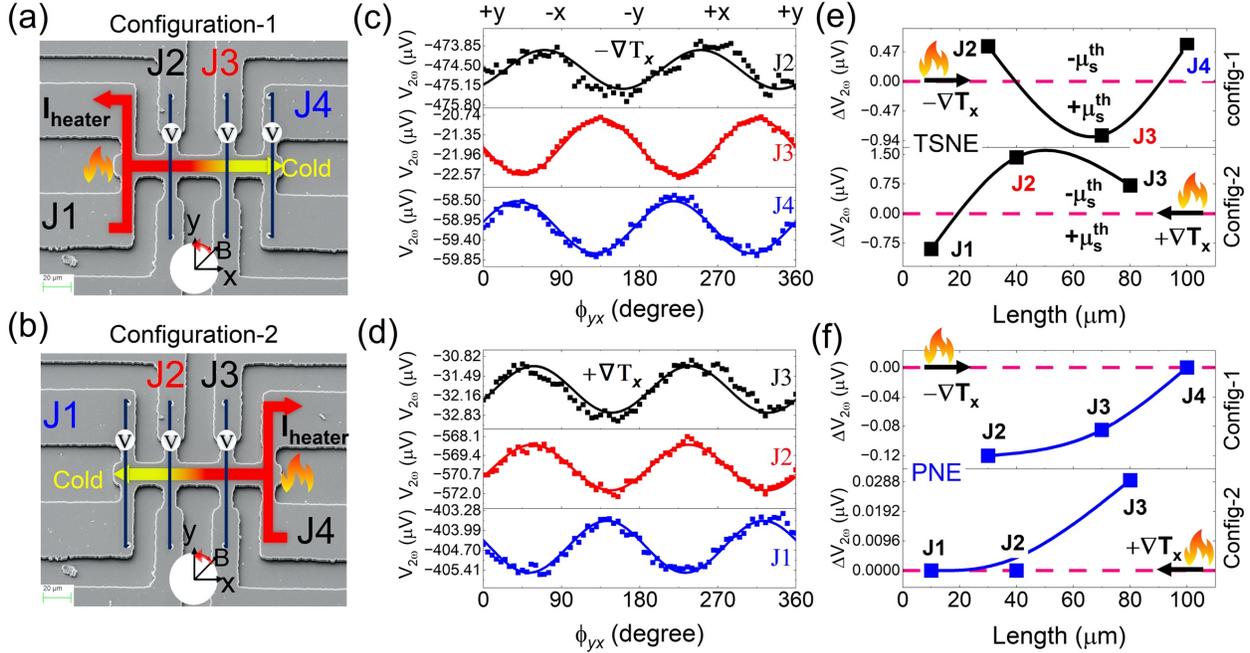

Figure 2. The scanning electron micrograph showing the representative device structure and experimental scheme for (a) configuration 1 and (b) configuration 2. The angle-dependent transverse magneto-thermoelectric response in sample 1 in the yx-plane at a constant applied magnetic field of 1 T (c) measured at J2, J3 and J4 in Configuration 1 and (d) measured at J3, J2 and J1 in Configuration 2. (e) The longitudinal distribution of the magnitude of transverse spin-Nernst effect for config 1 and config 2 with possible distribution of spin accumulation. (f) The planar Nernst effect response in control sample 2 showing the longitudinal distribution and response diminishes farther from the heat source. The solid line in (e) and (f) for representation only.

For the first experimental study, we fabricated a heterostructure composed of Py (25 nm)/MgO (1.8 nm)/SiO$_2$ (native)/p-Si (2 μm) (sample 1) as shown in Figure 2 (a,b), (Supplementary Section S1[11]). The resistivity of the Py layer is expected to be $4.92\times10^{-7}$ Ω-m[8]. The average charge carrier concentration in the p-Si[7] is expected to increase from ~$4\times10^{19}$ cm$^{-3}$ ($2.59\times10^{-5}$ Ω-m) (multiple control p-Si samples) to ~$5.45\times10^{19}$ cm$^{-3}$ ($1.9\times10^{-5}$ Ω-m) in case of sample 1. Hence, there is excess charge carrier concentration of~$1.45\times10^{19}$ cm$^{-3}$ in the sample 1, which leads to flexoelectronic polarization in the p-Si layer. The MgO layer is used to avoid any interfacial diffusion of the metal layer in Si layer. The strain gradient in the sample arises due to buckling of freestanding sample from the residual and processing thermal mismatch stresses including from material deposition[12,13], for which the maximum local strain could be as large as 4% based on previous studies[14]. It is noted that longitudinal inhomogeneities due to inhomogeneous mismatch stresses from geometrical and boundary imperfections may arise. However, these inhomogeneities will be average out due to bending and twisting deformation of the freestanding sample. In addition, the free charge carriers will quench any local inhomogeneities in the highly doped Si layer. We measured the angle dependent (yx-plane) transverse thermoelectric ($V_{2\omega}$ response) response in the sample 1 at 300 K and constant magnetic field of 1 T in two configurations: Configuration 1- heat junction J1 using a 2 mA 37 Hz of applied current and measure response at J2 (30 μm), J3 (70 μm), J4 (100 μm), and Configuration 2- heat junction J4 and measure response at J3 (30 mm), J2 (70 μm), and J1 (100 μm). The measured responses exhibit predominantly a $\sin 2\phi_{yx}$ angular symmetry (solid line fit) for both configurations as shown in Figure 2 (c,d) and Supplementary Table S1. The curve fit is found to be better for the responses farther from the heat source as shown in Figure 2 (c,d), which can potentially be attributed to larger thermal noise near the heat source. We do not observe $\sin\phi_{yx}$ symmetry in the angle dependent measurement corresponding to the transverse spin-Seebeck effect[15]. The offset $V_{2\omega}$ response is attributed to conventional thermopower, which arises due to asymmetries in the overall device structure, which will not be discussed in this study. The spatial distribution of the angle dependent thermoelectric response is shown in Figure 2 (e) for both configurations. From the spatial distribution of the response, we observe that the responses at junctions J3 and J4 are larger than that at junction J2 in the measurement configuration 1 even though junction J2 is nearer to the heating source as shown in Figure 2 (c,e) and expected to have larger thermal gradient. Similarly, the angle dependent thermoelectric responses measured at junctions J2 and J1 are larger than response measured at junction J3 even though it is nearer to the heat source in the measurement configuration 2 as shown in Figure 2 (d,e). Such a spatial distribution of the magneto-thermoelectric response is not feasible in the diffusive thermal transport regime since thermoelectric response should have decreased exponentially. Instead, the response at the farthest locations are higher than the nearest locations in both configurations. In addition, the measurements in both configurations show a sign reversal behavior in the measured responses.

The angular symmetry corresponding to $\sin 2\phi_{yx}$ can arise from the planar Nernst effect (PNE)[16] response from the ferromagnetic (Py) layer, which can be described as follows:

$$PNE \propto M \times (M \times \nabla T) \qquad\qquad (2)$$

where $M$ and $\nabla T$ are magnetization vector and temperature gradient, respectively. In this case, the sign of temperature gradient is constant and magnetization vector is aligned along the applied magnetic field for any give measurement configuration. Hence, the observed sign reversal cannot arise from the PNE behavior from the ferromagnetic material. Further, the thermal conductivity of 2 μm Si ($\kappa = \sim 80\ Wm/k$) [17] [18] is expected to be 4 times larger than that of the Py ($\kappa = \sim 20\ Wm/k$) [19] layer. The thermal resistance of the p-Si layer is expected to be ~320 times smaller than that of Py layer. As a consequence, the heat transport is across the Py layer will be negligible as compared to the p-Si layer in the plane of the sample and out of plane temperature difference will be insignificant. To further support our argument, we measured the angle dependent thermoelectric responses in a control Py/SiO$_2$ (25 nm)/p-Si sample (sample 2) as shown in Supplementary Figure S1 and Supplementary Section S2. The 25 nm of SiO$_2$ intermediate layer extinguishes the spin current as well as the interlayer flexoelectronic charge carrier transfer. In this sample 2, the symmetry and sign of the measured responses are same as expected for PNE behavior as shown in Figure 2 (f), which is consistent with that reported in the previous studies[16]. The PNE response in sample 2 is negative in configuration 1 whereas the response in sample 1 is positive at junction J2 and J4. Similarly, PNE response in sample 2 is positive in configuration 2 whereas response in sample 1 is negative at junction J1. The PNE response cannot reverse sign as already stated. Additionally, the PNE response decreases as expected for diffusive thermal transport as shown in Figure 2 (f). The magnitude of the angle dependent response in sample 2 is order of magnitude less than that in sample 1. From the symmetry[16], spatial distribution and magnitude of the angle dependent thermoelectric response, we eliminated PNE as the underlying cause of the observed behavior.

## B. Non-local resistance measurement

The flexoelectronic charge carrier transfer from metal to Si layer leaves the metal layer deficient in charge carrier. As a consequence, there is opposite flexoelectronic effect in the metal layer as well. Hence, the magneto-thermoelectric transport could arise in the Py layer itself instead of p-Si layer. In order to eliminate current leakage effect[20] and contribution of Py layer, we did angle dependent non-local resistance measurement on a Pt (15 nm)/MgO/p-Si (2 µm) sample (sample 3) as a function of temperature where dynamical multiferroicity is already demonstrated[7]. In sample 3, the resistivity of the Pt and p-Si layers are $2.52 \times 10^{-7}$ $\Omega$-m and $1.1 \times 10^{-5}$ $\Omega$-m[7], respectively. The expected charge carrier concentration in the p-Si layer is ~$9.4 \times 10^{19}$ cm$^{-3}$ due to flexoelectronic effect[7]. Hence, there is excess charge carrier concentration of~$5.4 \times 10^{19}$ cm$^{-3}$ in the sample 3, which leads to flexoelectronic polarization in the p-Si layer. In this second experiment, we applied a 2 mA, 37 Hz of current bias across the junction J1 in a Pt (15 nm)/MgO/p-Si sample as shown in Figure 3 (a). We, then, measured the angle dependent (yx-plane) non-local resistance across junctions J2 (30 µm), J3 (70 µm) and J4 (100 µm) for an applied magnetic field of 4 T as shown in Figure 3 (a). The measurement was done at 300 K, 100 K, 50 K, 25 K and 5 K. The measurement data for J2, J3 and J4 is shown in Figure 3 (b-d).

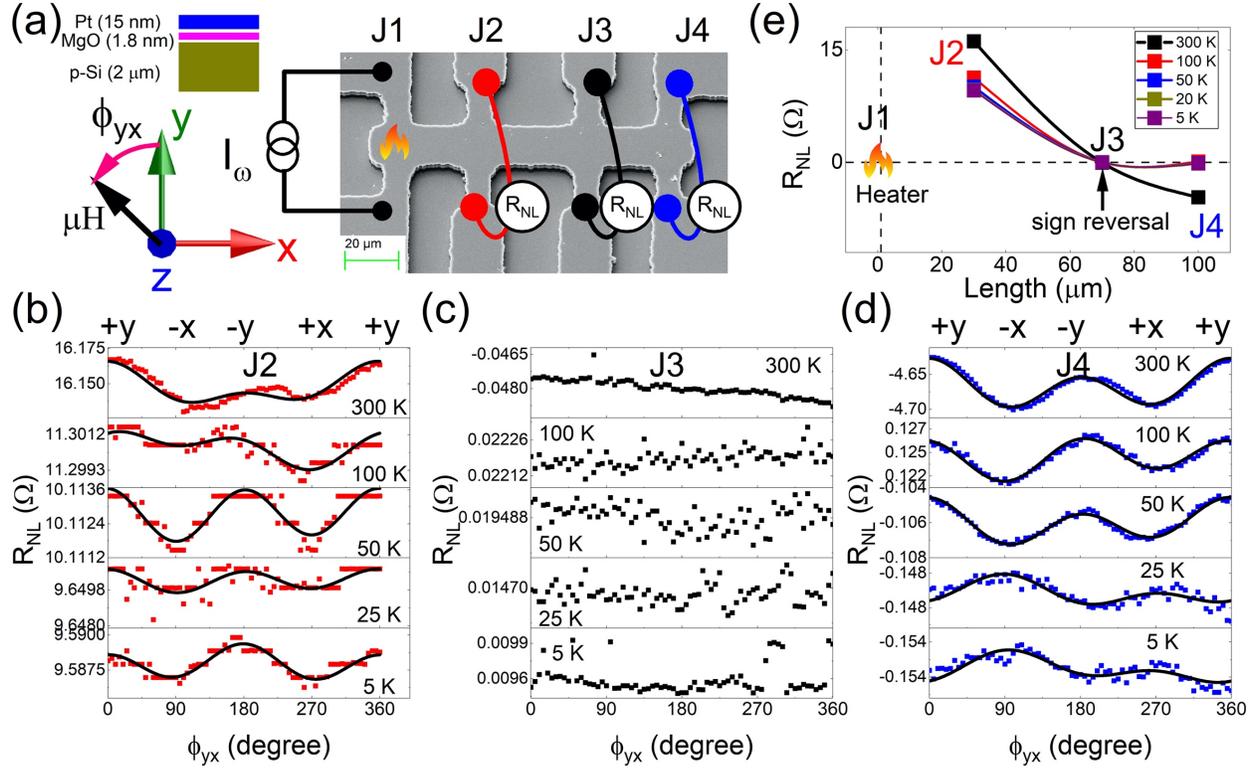

Figure 3. (a) A representative scanning electron micrograph showing the experimental scheme for angle dependent non-local resistance measurement where an alternating current bias is applied across junction J1 and responses are measured simultaneously across junctions J2, J3 and J4. The angle dependent non-local resistance at an applied magnetic field of 4 T in the yx-plane measured at (b) J2, (c) J3 and (d) J4 junctions and at 300 K, 100 K, 50 K, 25 K and 5 K. The black line represents the curve fit. (e) The spatial distribution of the amplitude of the responses at J2, J3 and J4 showing the sign reversal from potential distribution of temporal magnetic moment and spin accumulation.

The non-local resistance can be described using van der Pauw's theorem[20] as:

$$R_{NL} = R_{sq} e^{\frac{-\pi L}{w}} \qquad (3)$$

where $R_{sq} = \frac{\rho}{t}$, $\rho$, $t$, $L$ and $w$ are resistivity, thickness, length and width of channel. Based on the sample dimensions, the non-local resistance should be 5 m$\Omega$ at J2 and decrease exponentially along the length. Whereas, the measured non-local resistances are 16.17 $\Omega$, -0.048 $\Omega$ and -4.6 $\Omega$ at Hall junctions J2, J3 and J4, respectively, at 300 K as shown in Figure 3 (b-d). The non-local resistance measured at J2 is larger than even the longitudinal resistance (11.38 $\Omega$ between J2 and J3) of the sample at 300 K. In another study, the non-local resistance measurement on a p-Si sample is reported to be ~220 m$\Omega$ at 30 µm away and ~32 m$\Omega$ at 40 µm away from leakage current [21]. Hence, all the responses are orders of magnitude larger than that expected based on leakage current. Further, the response measured at junction J4 is much larger than that at J3 even though J4 is farther away from the current source and non-local response should be insignificant according to van der Pauw's theorem. Hence, leakage current can be eliminated as a source of the observed response.

The angle dependent behavior is expected to arise from relationship between magnetic field, flexoelectronic polarization and temporal magnetic moment. The peak modulations in the angle dependent response occurs when the magnetic field is aligned either parallel (+x) or anti-parallel (-x) to the direction of transport as shown in Figure 3 (b,d). Additionally, the magnitudes of the responses are different for parallel and anti-parallel configurations as shown in Figure 3 (b,d). Such a behavior can arise from chiral anomaly[22], which magnetoelectronic electromagnon are expected to exhibit since they are similar to photons. It is noted that chirality may arise due to twisting deformation of the sample from structural inhomogeneities. We do not observe any angular symmetry in the response measured at junction J3 as shown in Figure 3 (c), which indicates absence of spin polarization. In addition, a sign reversal, similar to magneto-thermoelectric response in sample 1, is observed in non-local response measured at both J3 and J4 as compared to response at J2 as shown in Figure 3 (e). In the absence of magnetic source, the observed angle dependent non-local resistance behavior cannot arise from any other known mechanism. Based on these experiments, the Py and Pt layers are not the underlying cause of magneto-thermoelectric and non-local resistance responses. These responses are expected to arise due to spin current in the p-Si layer in both the samples 1 and 3.

The magnitude of the non-local responses in sample 3 decreases as a function of temperature as shown in Figure 3 (b-e). The sign of the non-local responses also changes as the temperature is reduced as shown in Figure 3 (b-e). The conventional non-local resistance response cannot change sign as a function of temperature and only a spin dependent response can change sign. The measurement at J4 shows that the magnitude of the non-local resistance (offset response) reduces from 4.6 $\Omega$ at 300 K to ~0.125 $\Omega$ at 100 K, which is sharp reduction as shown in Figure 3 (d). In contrast, the non-local resistance measured at J2 decreases from 16.17 $\Omega$ at 300 K to 11.3 $\Omega$. This shows that the Hall junction nearest to the source shows the least reduction as compared to the farthest Hall junction as a function of temperature. The temperature dependent behavior suggests that the phonons are, most likely, to be the underlying cause of this non-local resistance behavior since they freeze at lower temperatures.

We eliminated the PNE and current leakage as an underlying cause of the magneto-thermoelectric response observed in sample 1. Hence, the transverse spin Nernst effect (TSNE) from the non-magnetic layer (p-Si in this study) [23-25] is expected to be the underlying cause of angle dependent magneto-thermoelectric response in sample 1, which can be described as:

$$TSNE \propto M \times (M \times \sigma_s^{th}) \qquad (4)$$

where $\sigma_s^{th}$ is the polarization vector of thermal spin accumulation. However, the thermal spin polarization vector in the p-Si layer has to be spatially modulated along the length of the sample in order to have the observed sign reversal. The electronic thermal transport under spin-orbit coupling (SOC) will give rise to transverse spin current of same polarity across the length of the sample without any sign reversal. Hence, the sign reversal cannot occur in electronic thermal transport under SOC, which eliminates conventional TSNE response as the underlying cause of the observed behavior.

In contrast, the spatially modulated thermal spin accumulation with opposite polarity can potentially arise from hypothesized magnetoelectronic electromagnon mediated process due to dynamical multiferroicity. We expect non-reciprocity ($\partial_t \boldsymbol{P}^+ \neq \partial_t \boldsymbol{P}^-$) due to lifting of degeneracy from broken structural inversion symmetry. As a consequence, the magnetoelectronic electromagnon with temporal magnetic moment $\mathbf{M}_t^{[1\bar{1}\bar{1}]}$ and $\mathbf{M}_t^{[1\bar{1}1]}$ will not cancel the $\mathbf{M}_t^{[\bar{1}11]}$ and $\mathbf{M}_t^{[\bar{1}1\bar{1}]}$ in the (110) cross-sectional plane[7] of the sample. Hence, the resulting thermal spin accumulation from superposition of magnetoelectronic electromagnon for the (110) cross-sectional plane can be described as:

$$\mu_s^{th}(x,y) \propto \boldsymbol{M}(x,y) \propto \sum \mathbf{M}_t^{<hlk>} = \mathbf{M}_t^{[1\bar{1}\bar{1}]} + \mathbf{M}_t^{[1\bar{1}1]} + \mathbf{M}_t^{[\bar{1}11]} + \mathbf{M}_t^{[\bar{1}1\bar{1}]} \qquad (5)$$

This superposition of the temporal magnetic moments of magnetoelectronic electromagnons, at any location, can give rise to either spin-up or spin-down spin accumulation depending upon their relative contribution as shown in Figure 1 (e). Therefore, the direction of the interlayer spin current as well as measured magneto-thermoelectric response will be a function of spatial coordinates having symmetry of TSNE behavior, which is expected to be the underlying cause of sign reversal observed in both the magneto-thermoelectric and non-local resistance responses. This can also explain absence of any angle dependent response in non-local resistance measurement at junction J3 as shown in Figure 3(c). Since, the measurements are undertaken over a finite Hall junction width and the measured response will be a summation over that finite width, which can be described as:

$$\pm \mu_s^{th}(measured) = \oint \mu_s^{th}(x,y) \qquad (6)$$

The resulting modulations in the local temporal magnetic moment can give rise to a smaller thermal spin accumulation and explains the observed spatial distribution of the measured magneto-thermoelectric and non-local resistance responses irrespective of the temperature gradient as shown in Figure 2 (e) and Figure 3 (e).

### C. Non-reciprocity of magnetoelectronic electromagnon

To support our mechanistic explanation, a direct evidence of broken structural inversion symmetry and flexoelectronic polarization mediated non-reciprocity ($\partial_t \boldsymbol{P}^+ \neq \partial_t \boldsymbol{P}^-$) is needed. Hence, we choose to measure the non-reciprocal transport response that arises from the structural inversion asymmetry in sample 3 for the third experiment. In the transport measurements, the resistance of the sample is a function of the direction of the current due to the non-reciprocity. As a consequence, the non-reciprocal resistance can be described as difference: $\Delta R = R(I) - R(-I)$. The non-reciprocal resistance can be described in terms of material properties as:

$$\Delta R \approx \frac{1}{2}\gamma R_0 IB \qquad (7)$$

where $R_0$, I, B and $\gamma$ are resistance at zero magnetic field, current, magnetic field and coefficient of magnetochiral anisotropy[26], respectively. The non-reciprocal transport behavior is studied using longitudinal second harmonic response since it is a quadratic function of applied ac bias as shown in equation 7.

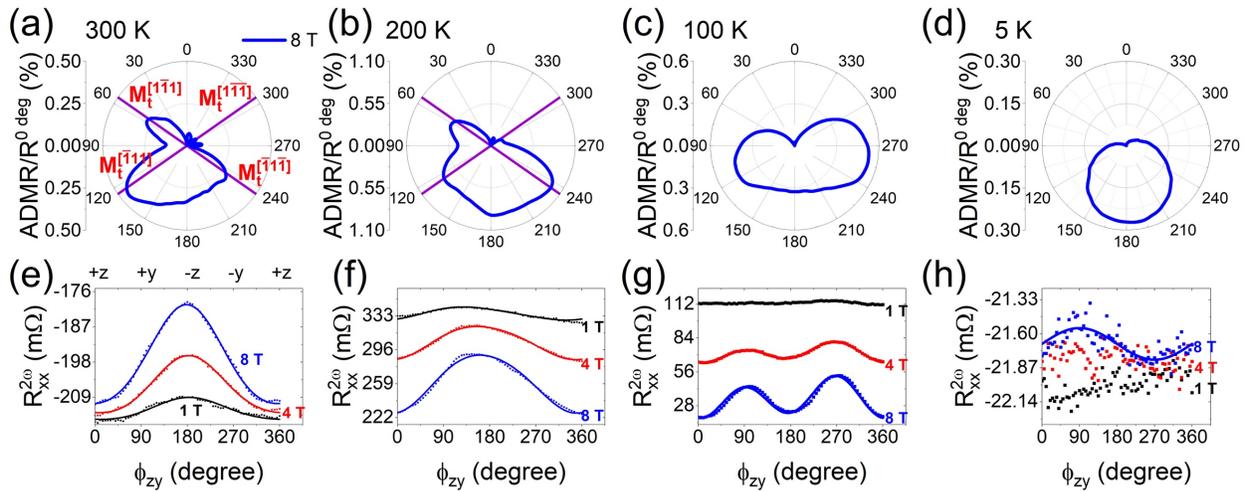

Figure 4. The angle dependent magnetoresistance behavior in sample 3 for the 8 T magnetic field at (a) 300 K, (b) 200 K, (c) 100 K and (d) 5 K. The angle dependent second harmonic behavior for applied magnetic field of 1 T, 4 T and 8 T in sample 3 at (e) 300 K, (f) 200 K, (g) 100 K and (h) 5 K. Solid line represent curve fit.

Hence, we measured the angle dependent longitudinal resistance and second harmonic resistance for an applied ac bias of 2 mA 37 hz and magnetic fields of 1 T, 4 T and 8 T in the sample 3 in this third experiment. The sample rotation was carried out in the zy-plane ($\phi_{zy}$) for a constant magnetic field, which was also the (110) cross-sectional plane. The angle-dependent MR (ADMR) measurements exhibits behavior corresponding to dynamical multiferroicity[7] as shown in Figure 4 (a,b). The magnitude of the ADMR shows peak corresponding to the <111> crystallographic directions[7] (at 54.7° from ±z-axis), which disappears at 100 K as shown in Figure 4 (a-d). When the large external magnetic field and, as a consequence, electron spins are aligned parallel to the temporal magnetic moment, the magnetoresistance reduces due to reduced scattering as compared to any other random direction. Hence, the ADMR response will have peaks corresponding to the temporal magnetic moments ($\boldsymbol{M}_t^{<111>} \propto \boldsymbol{P}_{F-El}^{<112>} \times \partial_t \boldsymbol{P}^{<110>}$) from dynamical multiferroicity. The ADMR response demonstrate temporal magnetic moments along $[1\bar{1}1]$, $[\bar{1}11]$ and $[\bar{1}1\bar{1}]$ crystallographic directions at both 300 K and 200 K. The relative magnitudes of the temporal magnetic moments are different at both temperatures. This shows that response along any location and for any given finite length will be a combination of relative contribution of temporal magnetic moments from dynamical multiferroicity as described by equation 5 and 6, which supports our mechanistic explanation for magnetoelectronic electromagnon mediated spin transport presented earlier. It is noted that the crystallography of temporal magnetic moment cannot be attributed to the Pt since it is polycrystalline. The dynamical multiferroicity disappears at lower temperatures, which is congruent to the reduction in the non-local resistance at junction J4 at 100 K and below in the second experiment on sample 3.

We, then, analyzed the angle dependent longitudinal second harmonic response at 300 K, 200 K, 100 K and 5 K, which demonstrate a clear cosine behavior due to non-reciprocal transport behavior at higher temperatures as shown in Figure 4 (e-f). The measured longitudinal second harmonic response diminished as the temperature was lowered to 100 K as shown in Figure 4 (g). At 5 K, the overall response was negligible and a weak sine behavior was observed at 8 T magnetic field only as shown in Figure 4 (h). Using the cosine fit, the coefficients of magnetochiral anisotropy are estimated to be 0.151 $A^{-1}T^{-1}$, 0.352 $A^{-1}T^{-1}$ and 0.071 $A^{-1}T^{-1}$ at 300 K, 200 K and 100 K, respectively. These values are similar to the reported value in Si FET interfaces (0.1 $A^{-1}T^{-1}$)[26-28] at 2.92 V of gate bias (except Si FET interfaces are two dimensional). While the raw magnitude of the non-reciprocal resistance is similar but coefficient of magnetochiral anisotropy is smaller than the maximum value of 1 $A^{-1}T^{-1}$ measured at 2 K reported in BiTeBr[27], which is one of the largest non-reciprocal response ever reported. However, the magnetochiral anisotropy in BiTeBr[27] drops to a value smaller than that in sample 3 at 200 K. This comparison demonstrates that the magnitude of the flexoelectronic polarization is large. However, a quantitative estimation of the flexoelectronic polarization is not feasible in this study. The conventional (electronic only processes) non-reciprocal responses increase[27] with reduction in temperature as opposed to the observed behavior in this study. The observation of dynamical multiferroicity in the ADMR response and large magnetochiral anisotropy in the non-reciprocal response along with their temperature dependent behaviors clearly demonstrate magnetoelectronic electromagnon (or magneto-active phonon) mediated behavior. The scattering from chiral spin fluctuations as described by Yokouchi et al[29] is expected to be the underlying mechanism, which are expected to arise from the magnetoelectronic electromagnons. Rest of the mechanisms[30] described in the literature are electronic and will not diminish at lower temperatures as stated earlier. Additionally, the observation of magnetochiral anisotropy clearly suggests that the magnetoelectronic electromagnon in flexoelectronically polarized p-Si are most likely to be chiral; potentially from twisting deformation. This also supports our previous contention of chiral anomaly in the angle dependent non-local resistance response but further studies are needed.

## D. Spatially modulated electronic property measurement

The magneto-thermoelectric and non-local resistance measurements showed a longitudinal sign reversal, which we attribute to spatially varying temporal magnetic moment from superposition of the magnetoelectronic electromagnon as described in equation 5 and 6. Since the magnetoelectronic electromagnon arise due to coupling of free charge carrier gradient and phonons, the spatial modulations in charge carrier concentration and magnetic moment must also arise according to equation 1, 5 and 6, which can be observed in Hall and anomalous Hall responses. Further, the equation 5 describes the superposition of magnetoelectronic electromagnons having opposite temporal magnetic moments and, possibly, chirality. The superposition of two magnetic states with opposite chirality[31] can be described as spin-density wave (SDW). Hence, the superposition of magnetoelectronic electromagnon is expected to also induce spatial (real space) modulations[32] in the spin density or incommensurate SDW in addition to other electronic properties as stated earlier.

Additionally, we need to establish existence of flexoelectronic effect or charge carrier transfer in the metal-semiconductor heterostructure. Currently, there is no standardized procedure to uncover this phenomenon since methods used for application of strain gradient inhibits magneto-transport measurements. However, strain state can be modulated using an applied charge current where differential thermal expansion from

Joule heating and resulting buckling will modify the strain gradient in a freestanding thin film sample as shown in Figure 5 (a) [8]. The electronic transport measurement as a function of current can be used to indirectly uncover the evidence of flexoelectronic effect. Hence a Hall resistance measurement as a function of space and applied current can be used to present an evidence of both flexoelectronic effect as well as real space modulations from magnetoelectronic electromagnon.

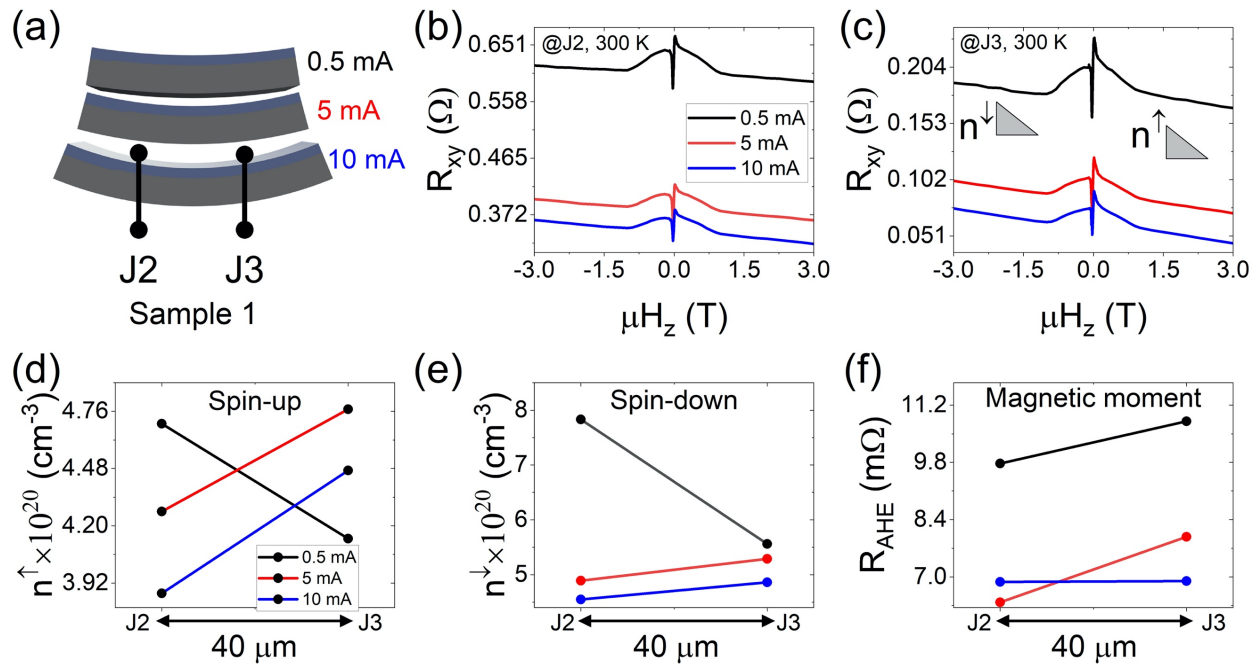

Figure 5. (a) schematic showing the increase in strain gradient from buckling of freestanding sample due to Joule heating from large longitudinal current bias. The Hall response measurement as a function of magnetic field swept from 3 T to -3 T at 300 K measured at Hall junction (b) J2 and (c) J3. The spatial modulation of the (d) spin-up charge carrier concentration, (e) spin-down charge carrier concentration, and (f) the anomalous Hall resistance as a function of current bias.

In the fourth experiment, we measured the Hall resistance response at Hall junctions J2 and J3 as a function of applied local current (0.5 mA, 5 mA and 10 mA at 37 hz) for an applied magnetic field sweep from 3 T to -3 T in the sample 1 as shown in Figure 5 (a). In the measurement, the overall transverse responses reduce as a function of applied current as shown in Figure 5 (b,c). The calculated values of the charge carrier concentrations and the anomalous Hall resistance are listed in the Supplementary Table S2. The Hall resistances are negative corresponding to electron as charge carrier. The larger strain gradient at higher current increases the flexoelectronic charge carrier transfer. Assuming the Fermi level to remain same, the measured average charge carrier concentration is expected to decrease as a function of current due to additional charge carrier transfer from Py to p-Si layer. The measured Hall resistances and average charge carrier concentrations indeed reduce as a function of current as shown in Figure 5(b-e), which prove our hypothesis. This a direct evidence of the flexoelectronic effect in our sample.

Further analysis show that the charge carrier concentration is larger at J2 as compared to J3 for both positive (spin-up) and negative (spin-down) magnetic fields as shown in Figure 5 (d,e) at 0.5 mA of applied current. For 5 mA and 10 mA of applied current, the charge carrier concentration behavior is reversed and the values at J3 are larger than that at J2 as shown in Figure 5 (d,e). The anomalous Hall resistance is larger at J3 (10.81 m$\Omega$) as compared to J2 (9.77 m$\Omega$) for 0.5 mA of current as shown in Figure 5 (f). The anomalous Hall resistance reduces with increased applied current. At 10 mA, the anomalous Hall resistances are similar at both locations (J2 (6.88 m$\Omega$) and J3 (6.9 m$\Omega$)). We estimated a temperature rise of ~70 K due to 10 mA of current. Using a control experiment at 375 K as shown in Supplementary Section S3 and Supplementary Figure S2, we eliminate uniform heating to be the underlying cause of change in charge carrier concentrations as well as anomalous Hall resistance. The evolution of the anomalous Hall resistance as well as charge carrier concentration as a function of location as well as current is attributed to the spatially modulations in the temporal magnetic moment, charge carrier concentration and spin density distribution from magnetoelectronic electromagnons, as hypothesized.

The negative Hall resistance in the sample 1 with 2 μm p-Si layer decreases as the current is increased in the previous experiment and any further increase in strain gradient should change the sign of the Hall resistance to positive. A similar behavior is expected to arise in case of anomalous Hall resistance as well. To induce larger strain gradient, we take a new sample (sample 4) for fifth experiment where thickness of the p-Si is reduced by chemical etching[33] to ~400 nm while the thickness of the Py layer is kept the same (Supplementary Section S1). Due to thinner p-Si layer, this new sample will have significantly larger strain gradient. In case of sample 4, the p-Si resistivity at 200 K in the heterostructure needs to be $6.72{\times}10^{-6}$ $\Omega$m as opposed to $5.45{\times}10^{-5}$ $\Omega$m[8]. The charge carrier concentration is expected to increase from $8.73{\times}10^{18}$ cm$^{-3}$ to ~$1.24{\times}10^{21}$ cm$^{-3}$,in case of sample 4, estimated based on previous report[8], which give rise to a large flexoelectronic polarization. As expected, we measure the positive Hall resistances at 200 K in this thinner sample for an applied local current of 2 mA 37 hz, as shown in Figure 6 (a), as compared to negative in the previous sample (sample 1 with 2 μm p-Si) due to larger flexoelectronic charge transfer as hypothesized in the Figure 1 (c). In addition, the sign of the anomalous Hall resistance also turns negative from positive. The concentration of the spin-down charge carrier decreases from $2.47\times10^{21}$ cm$^{-3}$ on the left Hall bar (J2) to $2.09\times10^{21}$ cm$^{-3}$ on the right Hall bar (J3) and concentration of spin-up charge carriers increases from $1.88\times10^{21}$ cm$^{-3}$ on the left to $2.17\times10^{21}$ cm$^{-3}$ on the right as shown in Figure 6 (b). The average charge carrier concentration decreases from $2.17\times10^{21}$ cm$^{-3}$ on the left to $2.13\times10^{21}$ cm$^{-3}$ on the right as shown in Figure 6 (b). Further, the anomalous Hall resistance is significantly smaller at J2 (-52.15 m$\Omega$) as compared to J3 (-77.46 m$\Omega$) as shown in Figure 6 (a). This reduction in the anomalous Hall resistance can be attributed to the net magnetic moment of opposite magnetoelectronic electromagnons aligning ferromagnetically/antiferromagnetically to the magnetic moment of Py layer. This reduction can also be attributed to larger local flexoelectronic charge transfer from Py layer. As compared to sample 1, we observe a crossover between spin-up and spin-down charge carrier concentrations along the length of the sample. The crossover in the spin dependent charge carrier concentration can be attributed to the longitudinal spin density modulation, which can, potentially, be considered as an incommensurate SDW from the superposition of magnetoelectronic electromagnons[31,34] as described in the equation 5 and stated earlier. This is a direct evidence of spin density, charge carrier concentrations and magnetic moment modulations (longitudinal) due to superposition of magnetoelectronic electromagnons. This behavior can also be called as inhomogeneous magnetoelectronic multiferroic effect since a spatially varying magnetic behavior give rise to spatially varying electronic properties. This is analogues to the inhomogeneous magnetoelectric effect where spatial magnetic inhomogeneity in a magnetic crystal[35-39] give rise to electric polarization and vice versa[40].

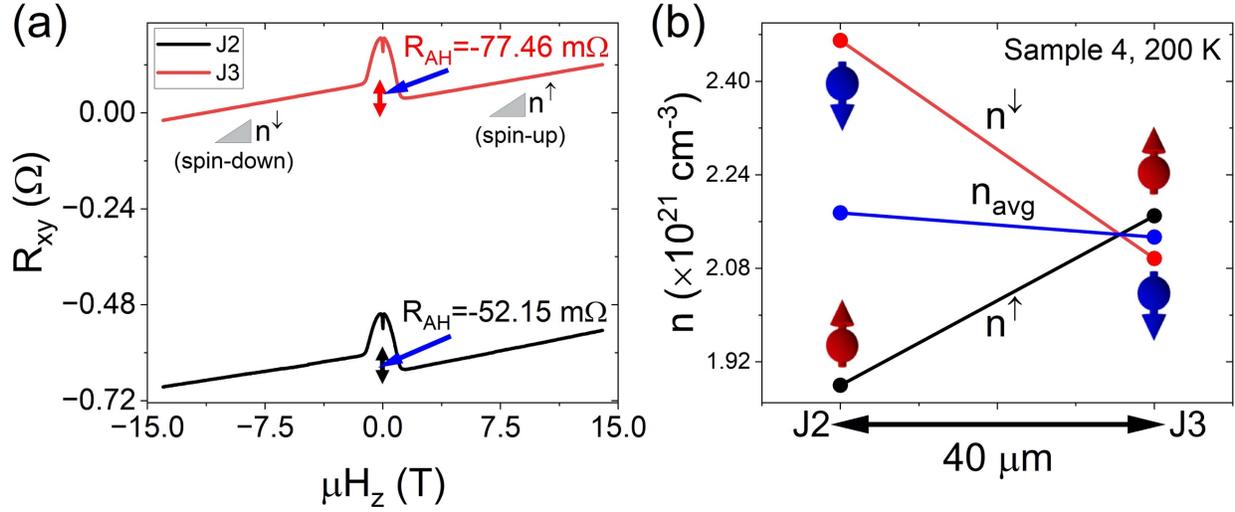

Figure 6. (a) Hall effect responses at junctions J2 and J3 in a Py(25 nm)/MgO/p-Si (~400 nm) sample measured at 200 K for an applied magnetic field from 14 T to -14 T, and (b) the longitudinal modulations in the charge carrier concentration and spin density showing incommensurate SDW like behavior.

### E. Edge dependent magnetoresistance measurement

Similar to longitudinal modulations, we expect charge carrier concentration, spin density and magnetic moment modulation along the width of the samples. To prove it, we measured the angle dependent longitudinal magnetoresistance and non-reciprocal second harmonic responses in a second Py/MgO/p-Si (2 μm) sample (sample 5) at two edges for an applied current bias of 2 mA 37 hz as shown in Figure 7 (a) in the sixth experiment. The charge carrier concentration in the p-Si layer is expected to increase to ~8.7×10$^{19}$ cm$^{-3}$ in case of sample 5 as opposed to 4×10$^{19}$ cm$^{-3}$. The measurement shows that the resistance measured at the right edge is ~23% smaller than that measured at the left edge as shown in Figure 7 (c,d). Similar behavior is observed in other samples too. The R$_{left}$=19.02 Ω and R$_{right}$=11.38 Ω are measured at 300 K in the sample 3 (with Pt).

These are attributed to the transverse charge carrier density modulations as hypothesized.

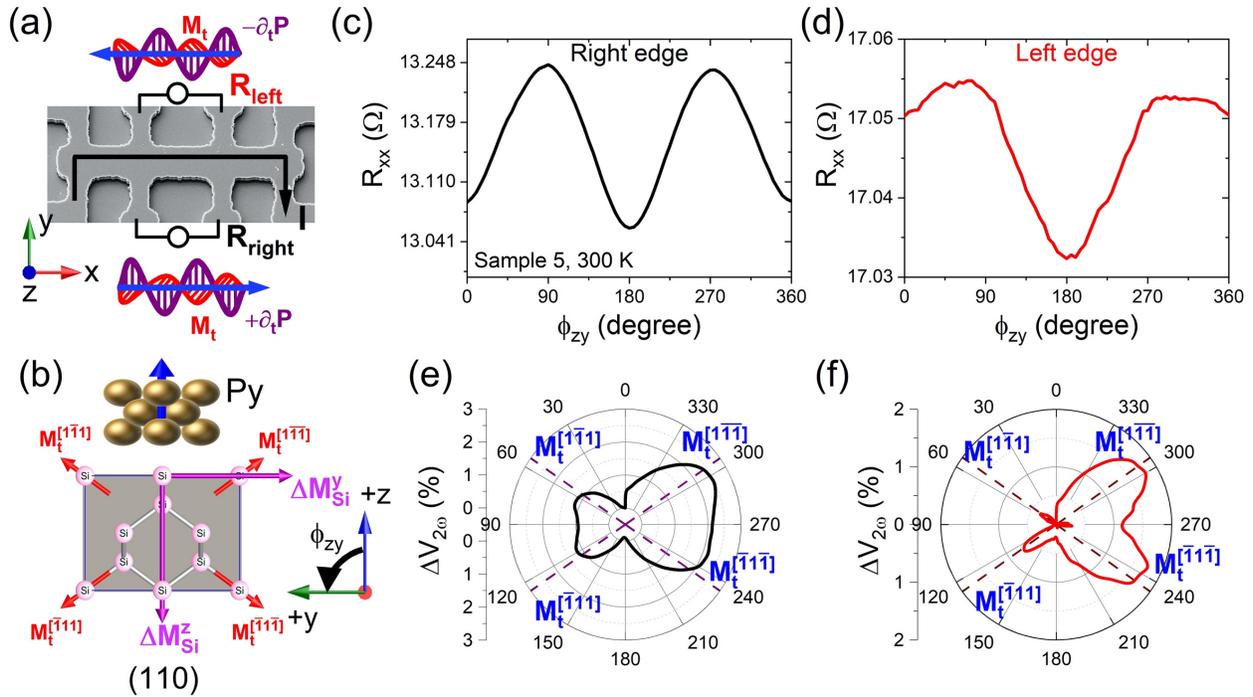

Figure 7. (a) A representative scanning electron micrograph showing the device structure and experimental scheme, (b) a schematic showing the temporal magnetic moments along <111> directions in the (110) cross-sectional plane leading to possible y and z-direction net magnetic moments. The angle dependent magnetoresistance in the second Py/MgO/p-Si sample measured at 300 K for an applied magnetic field of 8 T and angular rotation in the zy-plane (c) right side and (d) left side of the sample 5. The angle dependent longitudinal second harmonic non-reciprocal response measured at 300 K for an applied magnetic field of 8 T and angular rotation in the zy-plane (e) right side and (f) left side of the sample 5.

The ADMR response is composed of anisotropic magnetoresistance (AMR) ($\sin^2\theta_{zy}$) from Py layer and two responses having $\cos\theta_{zy}$ and $\sin\theta_{zy}$ symmetries (Supplementary Section S4). The giant magnetoresistance (GMR) response arises when in-plane current is applied in a thin film heterostructure where two ferromagnets are separated by a non-magnetic layer. A minimum in magnetoresistance occurs when the magnetic moment in both layers is aligned parallel whereas a maximum occurs when they are anti-parallel. In the sample 5, Py layer is one ferromagnet and the p-Si layer is effectively the second ferromagnet due to net magnetic moment from temporal magnetic moments. Hence, both $\cos\theta_{zy}$ and $\sin\theta_{zy}$ responses are attributed to the GMR behavior in sample 5. This measurement shows that a net magnetic moment (potentially canted) having component in the z-direction ($\Delta M_{Si}^{z}$) as well y-direction ($\Delta M_{Si}^{y}$) arises in the flexoelectronically polarized p-Si thin film as shown in Figure 7 (b). In addition, the net magnetic moments on the left and right edges are significantly different, which give rise to differences in measurements on left and right edge, respectively.

As compared to sample 3, where spin dependent scattering from temporal magnetic moment is observed in magnetoresistance, the second harmonic non-reciprocal responses show peaks corresponding to the temporal magnetic moments along <111> directions in the (110) cross-sectional plane of the Si thin film layer of the sample 5 as shown in Figure 7 (e,f,b). The response at the right edge corresponds to $\partial_t P^{[110]}$ and $\partial_t P^{[\bar{1}\bar{1}0]}$ as shown in Figure 7 (e) and can be described as:

$$M^{right} = \sum M_t^{right} = M_t^{[1\bar{1}\bar{1}]} + M_t^{[1\bar{1}1]} + M_t^{[\bar{1}11]} + M_t^{[\bar{1}1\bar{1}]} \qquad (8)$$

Whereas the response on the left edge arises primarily from $\partial_t P^{[110]}$ as shown in Figure 7 (f) and can be described as:

$$M^{left} = \sum M_t^{left} = M_t^{[1\bar{1}\bar{1}]} + M_t^{[\bar{1}11]} + M_t^{[\bar{1}1\bar{1}]} \qquad (9)$$

The difference in behavior on two edges is potentially due to spin-Hall effect of magnetoelectronic electromagnon. Hence, a superposition of temporal magnetic moments will give rise to different net magnetic moment (or spin accumulation) along two different edges. It is noted that the temporal magnetic moments are aligned along <111> directions that are at 54.7° from ±z- axis. As a consequence, the net magnetic moment will be canted from ±z- axis, which will give rise to components along z-axis and y-axis as shown in Figure 7 (b). This is the underlying cause of two GMR responses in the measured ADMR response as hypothesized earlier. Consequently, the out of plane GMR (z-direction- $\cos\theta_{zy}$) response is different between right (0.125%) and left (0.054%) edges as shown in Figure 7 (c,d), respectively. Similarly, the in-plane GMR (y-direction- $\sin\theta_{zy}$) responses are also estimated to be 0.026% and 0.0028% in the measurement on the right and left edges of the sample, respectively. Additionally, the AMR responses, which arises from Py layer only, in the sample are also modified due to spin current from the p-Si layer. The magnitudes of AMR response are estimated to be 1.27% and 0.056% in the measurement on the right and left edges of the sample, respectively.

We also measured the edge dependent magnetoresistance in sample 1 for an applied magnetic field from 3T to -3 T in the seventh experiment as shown in Figure 8 (a,b). In sample 1, the $R_{left}$=19.3 $\Omega$ and $R_{right}$=18.59 $\Omega$ are measured at 300 K at zero field. This shows transverse asymmetry similar to other samples. Most importantly, we observe a low field behavior with opposite polarity on opposite edges indicating spin-Hall effect of magnetoelectronic electromagnon. This give rise to the canted spin state and opposite net magnetic moment from superposition of temporal magnetic moments from non-reciprocal magnetoelectronic electromagnon. This canted spin state arises in Si layer and is a function of crystallographic orientation of the sample as shown previously[14] since temporal magnetic moment direction will be different along different crystallographic directions. The edge dependent experiments explicitly demonstrate the superposition of the magnetoelectronic electromagnons, which is the underlying cause of the spin dependent behavior observed in this work. Similar to previous study, the edge dependent experiments demonstrate inhomogeneity in the electronic properties, which is interpreted as inhomogeneous magnetoelectronic multiferroic effect.

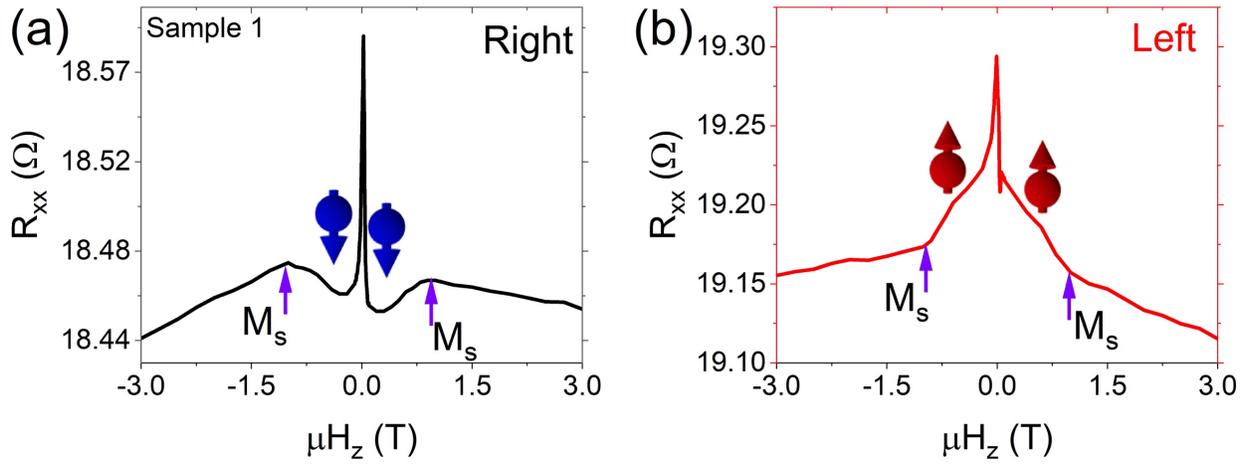

Figure 8. The edge dependent magnetoresistance measured for an applied magnetic field sweep from 3 T to -3 T (a) right edge and (b) left edge in the sample 1 at 300 K.

**F. Topological Nernst effect measurement**

The spin transport demonstrated in magneto-thermoelectric and non-local resistance measurements show very little dissipation. Further, the edge dependent measurement shows possible evidence of spin-Hall effect of magnetoelectronic electromagnon. Based on these results, we hypothesized that the magnetoelectronic electromagnon can potentially be topologically. The edge dependent behavior and magnetochiral anisotropy also suggested the same. Hence, we needed an experimental evidence of the topological Berry phase. In the freestanding sample, the longitudinal current lead to self-heating and a heat flow from the center of the sample to the boundaries and electrodes. As a consequence, the second harmonic Hall measurement as a function of out of plane magnetic field is expected to originate from the transverse thermoelectric responses (Nernst effects). We measured the second harmonic Hall response as a function of the out-of-plane magnetic field from 3 T to -3 T at an applied current of 5 mA in sample 1 in the eight experiment. The measurement at 300 K shows a transverse magneto-thermoelectric response attributed to the topological Nernst effect (TNE) [41] as shown in Figure 9 (a). The TNE response is significantly larger than anomalous Nernst effect (ANE) ($M_z \times \nabla T_x$) and ordinary Nernst effect (ONE) ($B_z \times \nabla T_x$) responses, which could not be discerned at 300 K as shown in Figure 9 (a) inset. Whereas, the second harmonic Hall response at 20 K is composed of distinct contributions from ONE, ANE and TNE responses as shown in Figure 9 (b). The TNE response is insignificant at 20 K as compared to 300 K, which is similar to the temperature dependent response observed in non-local resistance and non-reciprocal responses in sample 3. This result clearly supports the phononic (magnetoelectronic electromagnon) origin of the response from p-Si layer since thermal resistance of the p-Si layer is expected to be ~320 times smaller than Py layer, as stated earlier. This preliminary measurement shows that the magnetoelectronic electromagnon are potentially topologically, which leads to lack of dissipation observed in magneto-thermoelectric and non-local resistance measurements. This also explains previously reported large spin-Hall effect in Si[7,14]. This qualitative study only demonstrate observation and trend of TNE response and coefficients are not estimated. Further studies are needed to uncover the origin of the topological Berry phase as well as quantitative estimation of TNE response.

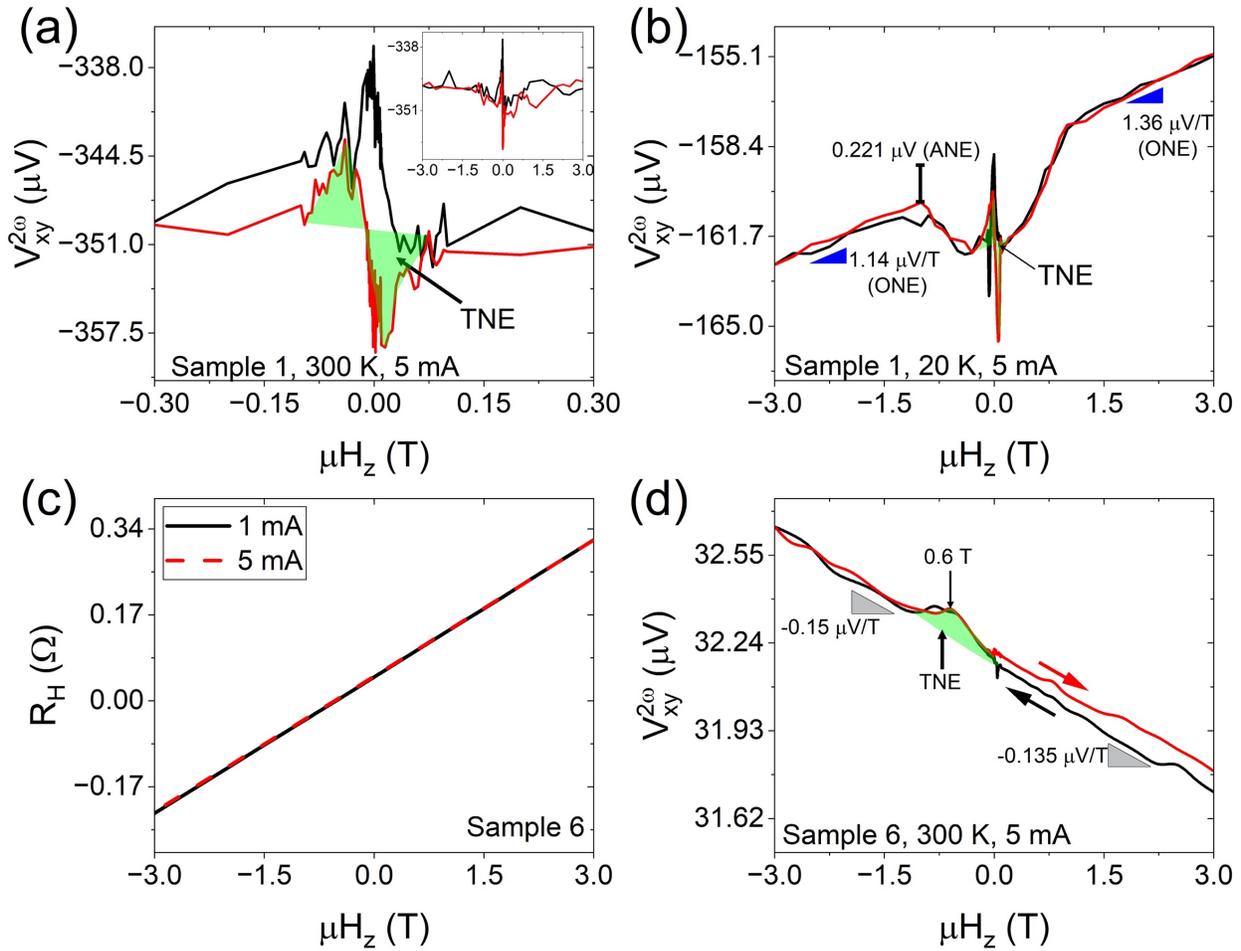

Figure 9. The second harmonic Hall response as a function of magnetic field from 3 T to -3 T for applied current of 5 mA in sample 1 (a) at 300 K and (b) at 20 K. (c) The Hall response measured at 300 K in sample 6 for 1 mA and 5 mA of applied current bias and a magnetic field sweep between 3 T and -3 T. (d) The second harmonic Hall response in sample 6 at 300 K for an applied current bias of 5 mA showing possible TNE behavior. The response in the green shaded region is expected to arise from topological Nernst effect.

To eliminate the contribution of Py layer, we, then, measured the Hall resistance and second harmonic Hall response in a MgO (2 nm)/p-Si (2 µm) sample (sample 6) at 300 K as a function of magnetic field from 3T to -3 T for an applied current of 1 mA and 5 mA in the ninth experiment. The Hall resistance and charge carrier concentration do not change when the current is increased from 1 mA ($3.48 \times 10^{19}$ cm$^{-3}$) to 5 mA ($3.47 \times 10^{19}$ cm$^{-3}$) as shown in Figure 9 (c) unlike the behavior observed in the fourth experiment in the sample 1 where the Hall response is a function of current as shown in Figure 5 (b,c). This presents an explicit proof of flexoelectronic charge carrier transfer in sample 1 and its absence in sample 6. We do not observe any behavior in the second harmonic Hall response at 1 mA of current as shown in Supplementary Figure S4. However, the second harmonic Hall response at 5 mA exhibits a negative slope and the behavior is attributed to the ONE due to the acoustic phonon scattering[42] as shown in Figure 9 (d). Furthermore, the magnitude of the slope is different for negative (-0.15 µV/T) and positive (-0.135 µV/T) magnetic fields, which indicated potentially a skew scattering behavior; supporting earlier observations. More importantly, a bump at the negative magnetic fields (green shaded region) is observed, as shown in Figure 9 (d). This is a manifestation of the topological Berry phase in thermal transport in the Si layer, leading to TNE[41]. An electromagnetic wave in an inhomogeneous medium experience a Berry gauge potential in the momentum space, which gives rise to the spin-Hall effect of light[43,44]. Bliokh and Freilikher[45] theoretically demonstrated that transverse acoustic waves in an inhomogeneous medium are analogous to electromagnetic waves. Hence, transverse acoustic waves will also experience phonon spin-orbit coupling due to the Berry gauge potential in the momentum space in a gradient index medium. The strain gradient can be considered equivalent to a gradient index medium in the freestanding thin films structure since a strain field will exist perpendicular to the phonon transport direction. The deflection of a phonon (ray) in an inhomogeneous medium is given by:

$$\delta \mathbf{r}_{tc} = -\sigma_c \lambdabar_{t0} \int_C \frac{\mathbf{p}_t \times d\mathbf{p}_t}{p_t^3} = -\sigma_c \lambdabar_{t0} \frac{\partial \Theta^B}{\partial \mathbf{p}_{tc}^{(0)}} \tag{10}$$

where $\sigma_c$, $\lambdabar_{t0}$, $\mathbf{p}_t$ and $\Theta^B$ are helicity, wavelength, momentum and Berry phase[45], respectively. As a consequence, the topological Berry phase is expected to arise in case of phonon transport in inhomogeneously strained Si thin films. The topological Berry phase of phonons when couples to flexoelectronic effect give rise to topological magnetoelectronic electromagnon quasiparticle excitations.

## III.  Discussion

Our magneto-thermoelectric and non-local resistance measurements show a large spin angular momentum transport at macroscopic (>100 μm) distance without any significant dissipation, which is the longest such distance ever reported. In case of quantum spin-Hall state[46], quantum-Hall antiferromagnet[47], antiferromagnetic insulators[48-50], magnetic insulators[51] and spin-superfluidity[52], the longest distance is an order of magnitude smaller than our measurement, at which spin transport is reported. The thermomagnetic effects, in general, can be described as cross product of temperature gradient with magnetic moment or magnetic field: $M \times \nabla T$, $B \times \nabla T$. There is no known thermomagnetic effect that will change sign while the direction of temperature gradient remaining same. As a consequence, the magneto-thermoelectric response and non-local resistance behavior cannot arise from thermomagnetic effects. Among all the spin dependent thermomagnetic effects, only the transverse spin-Seebeck effect give rise to a long-distance spin dependent response as well as sign reversal along the length of the sample[15,53]. The transverse spin-Seebeck effect arises due to the phonon driven spin distribution from the substrate[54,55]. The transverse spin-Seebeck effect is found to be absent in the suspended Py thin films[16] and our thin film samples are also freestanding. Further, the angle dependent response in sample 1 clearly eliminates transverse spin-Seebeck effect as the underlying reason of spin transport as already stated. The position dependent behavior is also contrary to the expected transverse spin-Seebeck effect response. The non-local resistance behavior observed in sample 3 also cannot arise from transverse spin Seebeck effect since there is no ferromagnetic spin source. Additionally, the experimental observation of dynamical multiferroicity and magnetochiral anisotropy shows that transverse spin-Seebeck effect is not the underlying cause of spin dependent behavior. Based on our experimental study, the superposition of the quasiparticle excitations called as magnetoelectronic electromagnon from dynamical multiferroicity is the underlying cause of the long-distance spin transport, magnetochiral anisotropy and spatial modulations in the charge carrier concentrations, spin density and magnetic moment. The magnetoelectronic electromagnons are most likely to be topological quasiparticle. The magnetoelectronic electromagnon can also provide a correct microscopic mechanism underlying the previously reported large spin-Hall effect[14], non-local transport[21] and magneto-thermal transport behavior[18].

## IV. Summary

In summary, we presented the experimental evidence of a quasiparticle excitations called as magnetoelectronic electromagnon in the degenerately doped p-Si based flexoelectronic heterostructures, which carries spin angular momentum and electronic charge. The magnetoelectronic electromagnons lead to long-distance spin transport; essential for the spintronics applications. The non-reciprocal magnetoelectronic electromagnon lead to large magnetochiral anisotropy at room temperature. Additionally, this work demonstrates superposition of topological magnetoelectronic electromagnons that give rise to spatial modulations in charge carrier density, spin density and magnetic moment. It is called as inhomogeneous magnetoelectronic multiferroic effect, which can give rise to incommensurate SDW like behavior. The magnetoelectronic electromagnon can also give rise to quantum interference and entanglement from opposite temporal magnetic moment (for example: $\mathbf{M}_t^{[1\bar{1}\bar{1}]}$ and $\mathbf{M}_t^{[\bar{1}11]}$) especially at room temperature. Hence, the magnetoelectronic electromagnon in flexoelectronic heterostructures can provide an alternate platform for rich and exotic high temperature material behavior that, traditionally, is not expected in conducting electronic systems.

**Author contributions**

AK, PCL and RGB have equal contribution to this work.

**Acknowledgement**

**Competing interests**

The authors declare no competing interests.